\documentclass[aps,prl,groupedaddress,twocolumn]{revtex4}
\usepackage{graphicx}
\usepackage{amsmath}
\usepackage{color}

\begin{document}

\title{Localization of Electromagnetic Fields in Disordered Fano Metamaterials}

\author{S. Savo}
\affiliation{Optoelectronics Research Centre, University of
Southampton, SO17 1BJ, UK}

\author{N. Papasimakis}
\affiliation{Optoelectronics Research Centre, University of
Southampton, SO17 1BJ, UK} \email{np3@orc.soton.ac.uk}

\author{N. I. Zheludev}
\affiliation{Optoelectronics Research Centre, University of
Southampton, SO17 1BJ, UK}

\date{\today}

\begin{abstract}
We present the first study of disorder in planar metamaterials consisting of strongly interacting metamolecules, where coupled electric dipole and magnetic dipole modes give rise to a Fano-type resonant response and show that positional disorder leads to light localization inherently linked to collective magnetic dipole excitations. We demonstrate that the magnetic excitation persists in disordered arrays and results in the formation of "magnetic hot-spots".
\end{abstract}

\maketitle
Recent years have seen a paradigm shift in studies of multiple scattering in random lattices: disorder instead of being an unwanted side effect, it is employed to enhance specific properties of materials \cite{genack11}. Particularly in electromagnetic systems, randomness starts to occupy a central position in numerous applications, where intriguing wave interference effects, such as Anderson localization \cite{genack00, segev07, wiersma11}, allow to confine light and enhance light-matter interactions with substantial implications for linear \cite{stockman94, shalaev99, LDOSmetalfilms1, LDOSmetalfilms2, giessen07, alu10} and non-linear electromagnetics \cite{nonlinear2,nonlinear1, segev07}, light emission \cite{randomlasers1,randomlasers2}, even cavity quantum electrodynamics \cite{QED}. To this end, cooperative and multiple scattering effects in metamaterials have also began to attract attention \cite{coherent, collapse, ccontrol,theory}. Indeed, metamaterials, owing to their unusual and exotic properties that range from cloaking and invisibility to negative refraction and super-lensing, allow for unprecedented control over electromagnetic wave propagation and hold promise of revolutionary future applications. However, although the importance of cooperative behaviour in metamaterials is being realized \cite{wegener11, sersic, theory}, studies of random metamaterials are rare and are typically concerned with the effects of disorder on the effective permittivity and permeability \cite{aydin04, gorkunov06,gollub09,helgert09}. In contrast to previous studies, here we investigate the role of electric and magnetic dipole excitations in Fano metamaterials with strongly interacting metamolecules and show that the localization is inherently linked to collective, subradiant (trapped) modes.
\begin{figure}
\includegraphics[width=85mm]{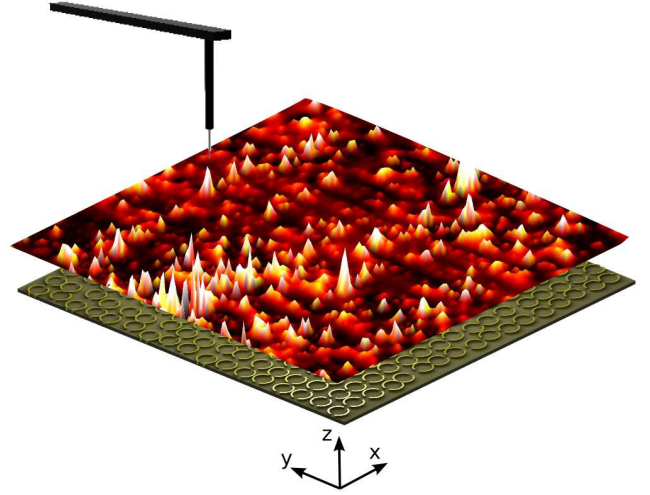} \caption{ Artistic impression of near-field fluctuations in a disordered metamaterial array. The enhanced electric field is collected by a monopole probe.}
\end{figure}

We study both regular and disordered metamaterial arrays consisting of metallic asymmetrically-split ring (ASR) resonators fabricated by standard lithographic techniques in $1.5$ mm thick FR4 dielectric substrates. The size of the unit cell is $15 \times 15~mm^2$, whereas the overall size of the arrays is limited to $15 \times 18$ unit cells. Disorder is introduced by displacing the center of each unit cell according to a random uniform distribution defined over the square interval $x\epsilon(-\alpha/2,\alpha/2)$, $y\epsilon(-\alpha/2,\alpha/2)$, where $\alpha=0.23~mm$. Disorder is quantified, by the degree of disorder, $D$, defined as the ratio of $\alpha$ over the unit cell side. The near field of the metamaterials is mapped by  a microwave near-field scanning microscope with hyperspectral capabilities. The scanning equipment is embedded in an anechoic chamber, where a broadband linearly polarized horn antenna placed $1.5~m$ from the planar array is used to illuminate the sample at normal incidence. An electric monopole oriented along the $z$ direction, mounted on top of a motorized stage is used to collect the near electric field (see Fig. 1). Near field maps have been obtained by probing the electric field in the direction normal to the sample (z-axis) with a spatial resolution of $2~mm$ in the sample plane and recording the amplitude and phase of the electric field by a vector network analyzer in the frequency range from $5~GHz$ to $7~GHz$. The experimental results are compared with full three-dimensional finite element calculations \cite{comsol} of finite size ($15$x$18$) regular and disordered metamaterial arrays. In the latter case, $10$ different realizations of disordered arrays for the same value of $D$ are considered. From such calculations the electric and magnetic dipole moments of each ASR are calculated according to the relations $\vec{p}=\frac{1}{i\omega}\int\vec{j}d^3r$ and $\vec{m}=\frac{1}{2c}\int\vec{r}\times\vec{j}d^3r$, respectively, and are compared in terms of their corresponding radiated powers according to $P_{el}=\frac{\omega^4}{12\pi \epsilon_0 c^3}|\vec{p}|^2$ and $P_{m}=\frac{\omega^4}{12\pi \epsilon_0 c^3}|\vec{m}|^2$.

The periodic array supports two types of coupled excitations forming a Fano-like system \cite{asrs}, an electric dipole mode (superradiant mode) where either the upper section or the bottom section of the split ring is excited (Figs. 2a \& 2c), and a magnetic dipole mode (trapped mode or subradiant mode) where antiphase currents flow in both arcs resulting in a magnetic dipole oscillating in the direction normal to the plane of the array (Fig. 2b). The magnetic excitation is weakly coupled to free-space and the mode appears trapped in the vicinity of the metamaterial surface \cite{asrs}. This minimal coupling, in turn, results in very strong interactions between metamolecules, providing ideal conditions to investigate localization and field enhancement effects in the metamaterial plane. In the far-field transmission spectrum (see Fig. 1d), the electric dipole mode results in a broad stopband extending throughout the measurement frequency range. The trapped mode manifests as a sharp roll-off, typical of Fano resonances, at about $5.6~GHz$, within the electric dipole band.

The near-field of a periodic array at the trapped-mode resonance ($5.67$ GHz) is shown in the experimental map of Fig.~3a, where as expected a regular field pattern can be seen throughout the metamaterial. Here, the excitation is quantified by the normalized electric field intensity, $I=|E_z|^2/<|E_z|^2>$, where $<|E_z|^2>$ is the average electric field intensity over the whole array. A slow modulation of the near field is attributed to finite size effects. The metamaterial response changes dramatically upon introducing disorder as shown in Fig.~3b. Here the near field of a disordered array with $D=0.15$ is presented at the same frequency ($5.67$ GHz). Although the displacement of the unit cells is relatively small ($15\%$ of the unit cell), the periodic pattern present in the regular case is much less pronounced, and a speckle pattern develops, where only certain clusters of  ASRs are strongly excited. 

The influence of disorder is reflected in the probability distribution function (pdf), $p(I)$, of the normalized electric field intensity $I$, defined as the ratio of the probability that the intensity of a metamolecule lies within a narrow interval $dI$ around $I$, over the interval $dI$. Regular arrays exhibit a narrow distribution with a finite width (see Fig. 3c) that originates both from imperfections of the experimental setup as well as the finite size of the sample. In contrast, disorder leads to a much broader, heavily skewed distribution and an exponential tail emerges at high intensities (Fig. 3d) which corresponds to the bright metamolecules of Fig. 3b and is not accessible in the case of regular arrays.
\begin{figure}
\includegraphics[width=75mm]{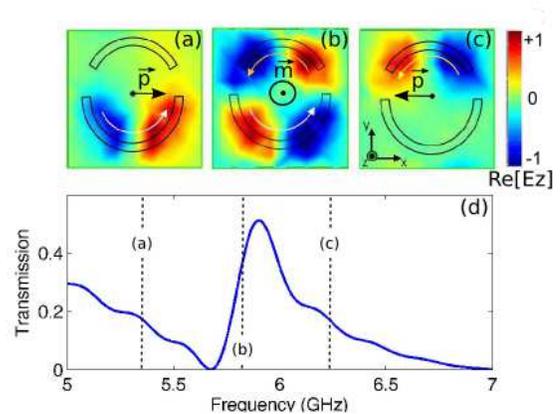} \caption{(a-c) Near field distribution of the real part of $E_{z}$ measured over a single unit cell of a regular ASR array at $5.3~GHz$ (a), $5.67~GHz$ (b) and $6.1~GHz$ (c). White and yellow arrows mark the direction of the currents flowing on the ASR arcs, where as black arrows represent the dominant dipole moment at each frequency, electric, $\vec{p}$, in (a) \& (c) and magnetic, $\vec{m}$, in (b). (d) Transmission of a regular ASR metamaterial array. Dashed lines mark the frequencies over which the near-field maps of (a-c) have been measured. In all cases, the electric field of the incident wave is oriented along the x-axis.}
\end{figure}

To investigate the link between the observed multiple scattering phenomena and the metamaterial response, we appeal to the results of the finite element simulations and decompose the response of each resonator in electric and magnetic dipole components. The radiating powers of the electric and magnetic dipole moments are presented as a function of the near field intensity in Figs.~3e \& 3f for regular and disordered arrays, respectively. Here, points at the low intensity part of the graph correspond to weakly excited metamolecules, while the "hot spots" populate the high intensity area. For regular arrays (Fig.~3e), the electric dipole response (black dots) is vanishing, whereas the magnetic dipole excitation is dominant (red squares). Furthermore, the magnetic dipole strength is positively correlated to the near field intensity, meaning that the more strongly excited metamolecules in the regular array are excited in the magnetic dipole mode. 

In the case of the disordered array (Fig. 3f), the situation is similar, although now the excitation of the metamolecules includes an electric dipole component that is no longer negligible.  However, in the moderate and high intensity region, it is evident that the ASRs with the more intense local fields, support predominantly magnetic dipole excitations. This indicates that the magnetic excitation of the ASRs persists in disordered arrays and leads to the formation of "magnetic hot-spots". Since such excitations are much stronger than the electric dipole ones, we argue that they constitute the main scattering mechanism.
\begin{figure}
\includegraphics[width=75mm]{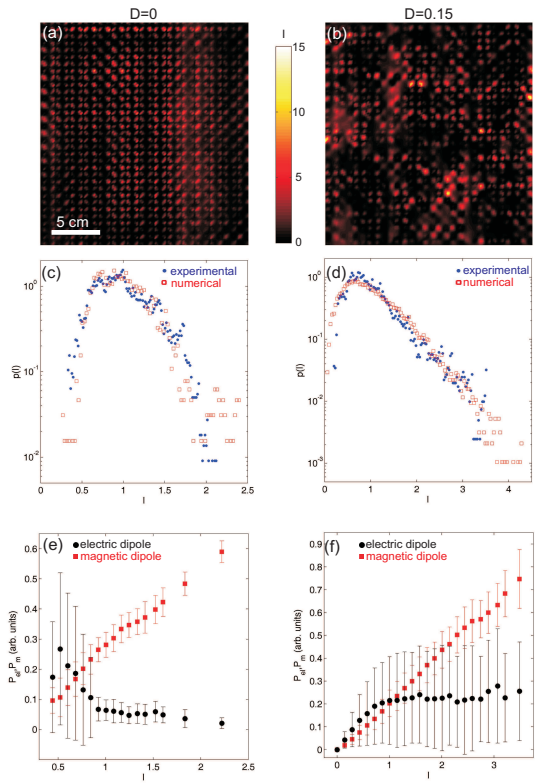} \caption{Characteristic experimental normalized near-field intensity maps of a regular ASR array (a) and a disordered array with disorder parameter $D=0.15$ (b). The corresponding probability density functions coarse-grained at the unit cell level over a $100$ MHz band around the trapped-mode resonant frequency are presented in panels (c) and (d). Blue full circles represent experimental measurements, while open red squares correspond to finite element simulations. The electric and magnetic dipole moments of each metamolecule derived from simulations are presented as a function of the normalized near-field intensity in (e) and (f) for regular and disordered arrays, respectively. Here, the dipole moment values have been grouped into intensity bins for presentation purposes. Error bars indicate the corresponding standard deviation of dipole moments within the intensity range of each bin.}
\end{figure}

The observed change in the statistics of the near-field intensity upon introducing disorder is a typical feature of multiple wave scattering in random media \cite{goodman85}. In the case of uncorrelated disorder and elastic isotropic scatterers, multiple scattering leads to a gaussian distribution of the real and imaginary part of the fields and the intensity follows the well known Rayleigh distribution, essentially a negative exponential. For rough or structured surfaces, such as random or fractal metal-dielectric composites, the situation is quite different, since multiple scattering of both propagating and standing waves takes place and there is no general form for the spatial intensity distribution. In addition, in the case of the planar metamaterials considered here, further complications arise from the fact that the scatterers (metamolecules) are strongly anisotropic and are both electrically and magnetically coupled, while dissipation losses take place in the dielectric substrate. Moreover, although the displacements of the metamolecules are random, their positions are still correlated and the disordered arrays partly retain characteristics of the regular lattice. Nevertheless, in spite of these complications, the observed intensity distributions agree qualitatively with theoretical and experimental results on multiple scattering of surface plasmon waves in semi-continuous metal-dielectric systems and nanostructured metallic films \cite{coello01,shalaev03,quelin06}. This is not surprising as in both cases localization and enhancement arises from interactions between the dipoles induced by the incident field either on the metamolecules or metallic clusters of the film. However, a crucial difference should be noted: in the ASR metamaterial case, the multiple scattering mechanism, i.e. the orientation and type (electric or magnetic) of the dipoles, as well as coupling to free-space is controlled by appropriate design of the metamaterial.

In conclusion, we report on near-field experimental investigations of metamaterials that support high-quality collective modes and demonstrate that disorder modifies the metamaterial response leading to intense localized excitations. Our results put forward metamaterials as an easily accessible, controlled environment to engineer the response of strongly interacting resonators to disorder. Finally, collective effects in disordered metamaterials allow the realization of arbitrary near-field landscapes which can be of interest for a number of applications, from site-specific sensing to optical trapping or even tailoring the radiation properties of light-emitting devices.

\begin{acknowledgments}
This work is supported by the Leverhulme Trust and the U.K.'s Engineering and Physical Sciences Research Council through the Metamaterials Programme Grant.
\end{acknowledgments}


\begin{thebibliography}{80}
\bibitem{genack11} J. Wang and A. Z. Genack, {\it Nature} {\bf 471}, 345 (2011).
\bibitem{genack00} A. A. Chabanov, M. Stoytchev, and A. Z. Genack,  {\it Nature} {\bf 404}, 850 (2000).
\bibitem{segev07} T. Schwartz, G. Bartal, S. Fishman, and M. Segev, {\it Nature} {\bf 446}, 52 (2007).
\bibitem{wiersma11} F Riboli et al, {\it Opt. Lett.} {\bf 36}, 127 (2011).
\bibitem{stockman94} M. I. Stockman, L. N. Pandey, L. S. Muratov, and T. F. George, {\it Phys. Rev. Lett.} {\bf 72}, 2486 (1994).
\bibitem{shalaev99} S. Gresillon et al., {\it Phys. Rev. Lett.} {\bf 82}, 4520 (1999).
\bibitem{LDOSmetalfilms1} K. Seal et al., {\it Phys. Rev. Lett.} {\bf 97}, 206103 (2006).
\bibitem{LDOSmetalfilms2} V. Krachmalnicoff, E. Castanie, Y. De Wilde, and R. Carminati, {\it Phys. Rev. Lett.}  {\bf 105}, 183901 (2010).
\bibitem{giessen07} D. Nau et al., {\it Phys. Rev. Lett.} {\bf 98}, 133902 (2007).
\bibitem{alu10} A. Alu and N. Engheta, {\it New J. Phys.} {\bf 12}, 013015 (2010).
\bibitem{nonlinear2} M. Breit et al., {\it Phys. Rev. B}  {\bf 64}, 125106 (2001).
\bibitem{nonlinear1} Sergey I. Bozhevolnyi, Jonas Beermann, and Victor Coello, {\it Phys. Rev. Lett.}  {\bf 90}, 197403 (2003).
\bibitem{randomlasers1} D. S. Wiersma, {\it Nature Phys.}  {\bf 4}, 359-367 (2008).
\bibitem{randomlasers2} C. Conti and A. Fratalocchi, {\it Nature Phys.}  {\bf 4}, 794-798 (2008).


\bibitem{QED} L. Sapienza et al., {\it Science} {\bf 327}, 1352-1355 (2010).
\bibitem{coherent} N. Papasimakis et al., {\it Phys. Rev. B}  {\bf 80}, 041102(R) (2009).
\bibitem{collapse} V. A. Fedotov et al, {\it Phys. Rev. Lett.}  {\bf 104}, 223901 (2010).
\bibitem{ccontrol} T. S. Kao, S. D. Jenkins, J. Ruostekoski, and N. I. Zheludev, {\it Phys. Rev. Lett.}  {\bf 106}, 085501 (2011).
\bibitem{theory} S. D. Jenkins and J. Ruostekoski, arxiv:1012.3928v1 (2010).

\bibitem{wegener11} M. Decker et al., {\it Phys. Rev. B} {\bf 84}, 085416 (2011).
\bibitem{sersic} I. Sersic, M. Frimmer, and A. F. Koenderick, {\it Phys. Rev. Lett.}  {\bf 103}, 213902 (2009).


\bibitem{aydin04} K. Aydin et al. {\it Opt. Express}  {\bf 12}, 5896 (2004).
\bibitem{gorkunov06} M. V. Gorkunov, S. A. Gredeskul, I. V. Shadrivov, and Y. S. Kivshar,{\it Phys. Rev. E} {\bf 73}, 056605 (2006).
\bibitem{gollub09} J. Gollub et al., {\it Appl. Phys. Lett.}  {\bf 91}, 162907 (2009).
\bibitem{helgert09} C. Helgert et al., {\it Phys. Rev. B}  {\bf 79}, 233107 (2009).
\bibitem{comsol} Finite element simulations were performed with the commercial solver COMSOL 3.4.
\bibitem{asrs} V. A. Fedotov et al., {\it Phys. Rev. Lett.}  {\bf 99}, 147401 (2007).
\bibitem{goodman85} j. W. Goodman, {\it Statistical Optics} (Wiley, New York, 1985).
\bibitem{coello01} S. I. Bozhevolnyi and V. Coello, {\it Phys. Rev. B}  {\bf 64}, 115414 (2001).
\bibitem{shalaev03} D. A. Genov, A. K. Sarychev, and V. M. Shalaev {\it Phys. Rev. E.} {\bf 67}, 056611 (2003).
\bibitem{quelin06} S. Buil, J. Aubineau, J. Laverdant, and X. Quelin, {\it J. Appl. Phys.}  {\bf 100}, 063530 (2006).


\end{thebibliography}
\end{document}